\documentstyle[preprint,aps,floats,epsfig,12pt]{revtex}

\begin{document}

\draft

\preprint{
\vbox{\halign{&##\hfil\cr
& ANL-HEP-PR-01-026  \cr
& January, 2002 \cr
&\vspace{0.6truein}   \cr
}}}

\title{Bottomonium Decay Matrix Elements from Lattice QCD with Two Light
Quarks}

\author{G.~T.~Bodwin and D.~K.~Sinclair}
\address{HEP Division,
Argonne National Laboratory, 9700 South Cass Avenue, Argonne, IL 60439}
\author{S.~Kim}
\address{Physics Department, Sejong University, Seoul 143-747, Korea}

\maketitle

\begin{abstract}
We calculate the long-distance matrix elements for the decays of the
$\Upsilon$ ($\eta_b$) and $\chi_b$ ($h_b$) states in lattice QCD with
two flavors of light dynamical quarks. We relate the lattice matrix
elements to their continuum counterparts through one-loop order in
perturbation theory. In the case of the leading $S$-wave matrix element,
we compare our result with a phenomenological value that we extract
from the experimental leptonic decay rate by using the theoretical
expression for the decay rate, accurate through relative order
$\alpha_s$. Whereas estimates of the leading $S$-wave matrix element
from quenched QCD are 40--45\% lower than the phenomenological value,
the two-flavor estimate of the same matrix element is close to the
phenomenological value. Extrapolating to the real world of $2+1$ light
flavors, we find that this matrix element is approximately 6\% higher
than the phenomenological value, but that the phenomenological value
lies within our error bars. We also compute the color-singlet and
color-octet matrix elements for $P$-wave decays. We find the value of
the color-singlet matrix element for $2+1$ flavors to be approximately
70\% larger than the quenched value and the value of the color-octet
matrix element for $2+1$ flavors to be approximately 40\% larger than
the quenched value.
\end{abstract}


\pagestyle{plain}
\parskip 5pt
\parindent 0.5in

\section{Introduction}

Bottomonium is a nonrelativistic system: the velocity $v$ of the $b$ and
$\bar{b}$ quarks in the center-of-mass frame is much less than unity
($v^2 \approx 0.1$). Bodwin, Braaten and Lepage \cite{bbl} have shown
that, within the framework of Nonrelativistic Quantum Chromodynamics
(NRQCD), the smallness of $v$ allows one to expand the decay rates into
light hadrons and/or electromagnetic decay products in powers of $v$.
Each term in this velocity expansion can be expressed as a finite number
of terms, each of which is a product of a long-distance ($\sim 1/M_b v$)
matrix element of a four-fermion operator between bottomonium states and
a short-distance ($\sim 1/M_b$) parton-level decay rate. Owing to the
asymptotic freedom of QCD, the short-distance parton-level decay rate
can be calculated perturbatively.

The $S$-wave bottomonium decay rates can be expressed, through
next-to-leading order in $v^2$, as
\begin{eqnarray}
\label{eqn:swave}
\Gamma({}^{2s+1}S_{2s+1}\rightarrow X) & = & 
{\cal G}_1({}^{2s+1}S_{2s+1})\,2\, {\rm Im\,}f_1
({}^{2s+1}S_{2s+1})/M_b^2 \nonumber \\
& + & {\cal F}_1({}^{2s+1}S_{2s+1})\,2\,{\rm Im\,}g_1
({}^{2s+1}S_{2s+1})/M_b^4\,.
\end{eqnarray}
Similarly the $P$-wave bottomonium decay rates at lowest non-trivial
order in $v$ are given by
\begin{eqnarray}
\label{eqn:pwave}
\Gamma({}^{2s+1}P_J\rightarrow X) & = & {\cal H}_1({}^{2s+1}P_J)
\,2\,{\rm Im\,}f_1 ({}^{2s+1}P_J)/M_b^4\nonumber \\
& + & {\cal H}_8({}^{2s+1}P_J)\,2\,{\rm Im\,}f_8
({}^{2s+1}S_{2s+1})/M_b^2\,.
\end{eqnarray}
The $f$'s and $g$'s are proportional to the short-distance rates for the
annihilation of a $b\bar{b}$ pair from the indicated ${}^{2s+1}L_J$
state, while ${\cal G}_1$, ${\cal F}_1$, ${\cal H}_1$, and ${\cal H}_8$
are the long-distance matrix elements.\footnote{Our quantities ${\cal
H}_1$ and ${\cal H}_8$ are related to the quantities $H_1$ and $H_8$ in
Ref.~\cite{Bodwin:1992ye} by ${\cal H}_1=M_b^4 H_1$ and ${\cal
H}_8=M_b^2 H_8$.} The subscripts $1$ and $8$ indicate that the
$b\bar{b}$ pair is in a relative color-singlet or color-octet state.
If one works to leading order in $v$ in the NRQCD Lagrangian, then the
matrix elements of the spin-singlet and spin-triplet states are equal.

In earlier papers \cite{bks-1}, we reported lattice NRQCD calculations
of ${\cal G}_1$, ${\cal F}_1$, ${\cal H}_1$, and ${\cal H}_8$ for the
$\Upsilon$ ($\eta_b$) and $\chi_b$ ($h_b$) states that made use of
quenched gauge-field configurations with inverse lattice spacings
$a^{-1} \approx 2.4~{\rm GeV}$ and $a^{-1} \approx 1.37~{\rm GeV}$. We
found that the value of the best-measured matrix element ${\cal G}_1$ is
40--45\% below a phenomenological value that we extracted from the
leptonic width of the $\Upsilon$ and the theoretical expression for the
width, accurate through relative order $\alpha_s$.\footnote{The
phenomenological value that we quote in the present paper is based on a
slightly different value for $\alpha_s$ than was used in
Ref.~\cite{bks-1}.} The NRQCD Collaboration \cite{Davies:1994mp,lepage}
had noted that at least part of the discrepancy is likely due to the use
of the quenched approximation. The reason that the quenched
approximation underestimates the matrix element is that the distance
scale associated with the bottomonium bound state [order $1/(M_b v)$] is
considerably larger than the scale at which the matrix elements sample
the wave function (order $a$, which is order $1/M_b$). If we fix the
lattice QCD coupling at $1/(M_b v)$ to a value that yields good
agreement with the bottomonium spectrum, then, in the quenched
approximation, the coupling at $a$ will be weaker than it should be.
Hence, the wave function at the origin will be too small, leading to a
prediction for the bottomonium decay rate that is too small.

In this paper we present calculations of the decay matrix elements for
the $\Upsilon$ ($\eta_b$) and $\chi_b$ ($h_b$) states that make use of
gauge configurations containing the effects of 2 flavors of light
dynamical (staggered) quarks. These calculations confirm that most, if
not all, of the discrepancy in the previous calculations of the matrix
elements was, in fact, due to quenching. Our results, when extrapolated
to three light flavors, lead to a slight overestimate of the $\Upsilon$
decay rate.

The remainder of this paper is organized as follows. In Sec.~II we define
the required matrix elements in the continuum and on the lattice and
describe the lattice implementation of NRQCD that we use in our
calculations. Section~III contains an outline of the perturbative
calculation that we use to relate the lattice matrix elements to their
continuum counterparts. We present our results in Sec.~IV, and Sec.~V
contains our conclusions.

\section{Matrix elements and lattice NRQCD}

In the leading non-trivial order in $v$, the NRQCD Lagrangian for the
bottom quark and antiquark is
\begin{equation}
{\cal L}_B = \psi^\dag \left( D_t - {{\bf D}^2 \over 2 M_b} \right)\psi
           + \chi^\dag \left( D_t + {{\bf D}^2 \over 2 M_b} 
\right)\chi\,,
\label{lagrangian}
\end{equation}
where $\psi$ is the quark annihilation operator and $\chi$ is the
antiquark creation operator. $D_t$ and {\bf D} are the gauge-covariant
temporal and spatial derivatives. Note that, although one can obtain the
correct leading-order spectroscopy in the Coulomb gauge by replacing
{\bf D} with the simple (non-covariant) gradient operator, the covariant
operator is needed to calculate the octet $P$-wave decay matrix element,
even at lowest non-trivial order. Not surprisingly,
Eq.~(\ref{lagrangian}) is just the Euclidean-time Schr\"{o}dinger
Lagrangian for the bottom quarks and antiquarks.

We work to leading order in $v$ in the Lagrangian. As we have mentioned,
at this order, the matrix elements of the spin-triplet and spin-singlet
states are identical. Therefore, we approximate all of the long-distance
matrix elements in Eqs.~(\ref{eqn:swave}) and (\ref{eqn:pwave}) as
spin-singlet matrix elements. Using the leading-order Lagrangian, we are
able to compute the order-$v^2$ $S$-wave matrix element ${\cal F}_1$,
with an error of order $v^4$. Note, however, that, in order to obtain a
full relative-order-$v^2$ computation of the $S$-wave decay rate, we
would need to compute ${\cal G}_1$ through relative-order $v^2$. This
would require relative-order-$v^2$ terms in the Lagrangian, in which
case the spin-singlet and spin-triplet states would be distinguished.

In terms of the fields $\chi$ and $\psi$, the spin-singlet matrix
elements that we compute are
\begin{mathletters}
\begin{eqnarray}
{\cal G}_1 & = & \langle {}^1S_0|\psi^{\dag}\chi\chi^{\dag}\psi
|{}^1S_0\rangle 
\label{eqn:g1}\,, \\
{\cal F}_1 & = & \langle {}^1S_0|\psi^{\dag}\chi\chi^{\dag} 
(-{\textstyle \frac{i}{2}}\stackrel{\leftrightarrow}{\bf D})^2\psi 
|{}^1S_0\rangle\,, 
\label{eqn:f1} \\
{\cal H}_1 & = &\langle {}^1P_1|\psi^{\dag}(i/2)
\stackrel{\leftrightarrow}{\bf D}
\chi\cdot\chi^{\dag}(i/2)\stackrel{\leftrightarrow}{\bf D}\psi
|{}^1P_1\rangle\,, 
\label{eqn:h1} \\
{\cal H}_8 & = & \langle {}^1P_1|\psi^{\dag}T^a\chi\chi^{\dag}T^a\psi
|{}^1P_1\rangle\,,
\label{eqn:h8}
\end{eqnarray}                                                   
\label{matrix-elements}%
\end{mathletters}%
where $\chi^{\dag}\stackrel{\leftrightarrow}{\bf D}\psi \equiv
\chi^{\dag}{\bf D}\psi-({\bf D}\chi)^{\dag}\psi$. 

The vacuum-saturation approximation is valid for the color-singlet
matrix elements and is accurate up to errors of relative order $v^4$
(Ref.~\cite{bbl}). In that approximation, Eqs.~(\ref{eqn:g1}),
(\ref{eqn:f1}), and
(\ref{eqn:h1}) become
\begin{mathletters}
\begin{eqnarray}
{\cal G}_1 & \approx & {\cal G}_1^{\rm VS}=
\langle {}^1S_0|\psi^{\dag}\chi|0\rangle
                       \langle 0|\chi^{\dag}\psi|{}^1S_0\rangle\,, \\
{\cal F}_1 & \approx & {\cal F}_1^{\rm VS}=
\langle {}^1S_0|\psi^{\dag}\chi|0\rangle
\langle 0|\chi^{\dag}(-{\textstyle\frac{i}{2}}
\stackrel{\leftrightarrow}{\bf D})^2
\psi |{}^1S_0\rangle\,, \\
{\cal H}_1 & \approx &{\cal H}_1^{\rm VS}=
\langle {}^1P_1|\psi^{\dag}(i/2)
\stackrel{\leftrightarrow}
{\bf D}\chi|0\rangle\cdot\langle 0|\chi^{\dag}(i/2)
\stackrel{\leftrightarrow}
{\bf D}\psi|{}^1P_1\rangle\,.
\end{eqnarray}
\end{mathletters}%
One can express vacuum-saturation values of the color-singlet matrix
elements as ${\cal G}_1^{\rm VS}= \frac{3}{2\pi}|R_S(0)|^2$ and ${\cal
H}_1^{\rm VS}= \frac{9}{2\pi}|R'_P(0)|^2$, where $R_S(0)$ is the
radial wave function of the $S$-wave state at the origin and $R'_P(0)$ is
the derivative of the radial $P$-wave wave function at the origin
\cite{bbl}. These are the quantities that appear in decay rates in the
color-singlet model. In contrast, the term proportional to ${\cal H}_8$
is absent in decay rates in the color-singlet model. ${\cal H}_8$ is the
probability of finding a $b\bar{b}g$ component in $P$-wave bottomonium,
with the $b\bar{b}$ in a color-octet state.

In our lattice calculation of these matrix elements, we transform our
gauge field configurations to the Coulomb gauge. For this gauge choice,
we can replace the covariant ${\bf D}$ with the non-covariant ${\bf
\nabla}$ in Eq.~(\ref{matrix-elements}). Corrections to this
replacement are suppressed by $v^2$.

We employ various discretizations of the derivative operator. For the
operator ${\cal H}_1$, we replace the covariant derivative ${\bf D}$ with
the non-covariant finite difference
${\bf \delta}$, which is defined by
\begin{equation}
\delta_i\psi(x) = \frac{1}{2}[\psi(x+{\bf i}) - \psi(x-{\bf i})]\,,
\end{equation}
where ${\bf i}$ is the unit vector in the $i$th spatial direction. For
${\cal F}_1$ we employ four different discretizations of ${\bf D}^2$:
\begin{mathletters}
\begin{eqnarray}
\Delta^{(2)}({\rm non})\psi(x) &=& \sum_i[\psi(x+{\bf i})+\psi(x-{\bf i}) 
- 2\psi(x)]\,,       \\
\Delta^{(2)}({\rm cov})\psi(x) &=& \sum_i\left\{\frac{1}{u_0}
[U_i(x)\psi(x+{\bf i}) + 
U_i^\dag(x-{\bf i})\psi(x-{\bf i})] - 2\psi(x)]\right\}\,,
\label{cov-2}
\end{eqnarray}
\begin{eqnarray}
\psi^\dag\Delta^{(2)}({\rm non}_2)\chi &=& -\sum_i[(\delta_i\psi)^\dag
\delta_i\chi]\,, \\
\psi^\dag\Delta^{(2)}({\rm cov}_2)\chi &=& 
-\sum_i[(d_i\psi)^\dag d_i\chi]\,,
\end{eqnarray}
\label{disc-delsq}
\end{mathletters}
where the covariant finite difference $d$ is defined by
\begin{equation}
d_i\psi(x) = \frac{1}{2u_0}[U_i(x)\psi(x+{\bf i}) - 
                            U_i^\dag(x-{\bf i})\psi(x-{\bf i})]\,,
\end{equation}
and $u_0$ is the tadpole contribution to $U$. We adopt the definition
$u_0=\langle\frac{1}{3}U_{\rm plaq}\rangle^{1/4}$.

\begin{figure}[htb]
\centerline{ 
\begin{picture}(200,125)
\put(50,95){\oval(80,40)}
\put(140,95){\oval(100,40)}
\put(0,65){\line(1,0){200}}
\put(47.5,35){\oval(75,40)}
\put(142.5,35){\oval(95,40)}
\put(10,95){\circle*{10}}
\put(190,95){\circle*{10}}
\put(10,35){\circle*{10}}
\put(190,35){\circle*{10}}
\put(90,95){\circle*{5}}
\put(85,35){\circle*{5}}
\put(95,35){\circle*{5}}
\put(45,95){\vector(-1,0){30}}
\put(55,95){\vector(1,0){32.5}}
\put(135,95){\vector(-1,0){42.5}}
\put(145,95){\vector(1,0){40}}
\put(47,92){$T$}
\put(137,92){$T'$}
%\put(195,10){,}
\end{picture}}
\caption{Lattice calculation of a matrix element of a four-fermion
operator. The large discs represent the sources and sinks; the smaller
discs represent the four-fermion and point source operators. The lines
are the nonrelativistic quark propagators.}
\label{fig:matrix}
\end{figure}
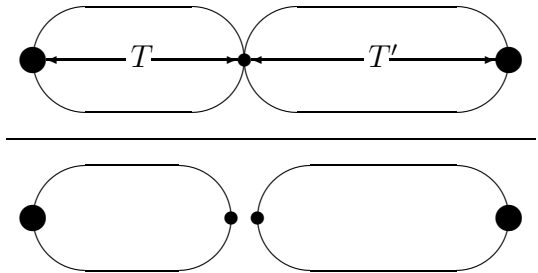

On the lattice, we obtain such matrix elements by measuring the
expectation value in the gluon background of a product of three
operators: a source for a $b\bar{b}$ pair with the appropriate quantum
numbers at a (Euclidean) time $-T$, the appropriate four-fermion
operator at time zero, and a sink for the $b\bar{b}$ pair at time $T'$.
For convenience, and in order to reduce noise, we divide this
expectation value by the product of two other expectation values. One is
the expectation value of the product of the numerator source for the
$b\bar{b}$ pair at time $-T$ and a point sink that annihilates the
$b\bar{b}$ pair at time zero;  the other is the expectation value of the
product of a point source that creates a $b\bar{b}$ pair at time zero
and the numerator sink, which annihilates the $b\bar{b}$ pair at time
$T'$. This ratio is illustrated in Fig.~\ref{fig:matrix}.
\narrowtext 
\widetext

In the cases of ${\cal G}_1$ and ${\cal H}_1$, this ratio approaches the
ratio of the matrix element to its vacuum-saturation approximation in
the limit $T,T' \rightarrow \infty$. Hence, it gives an indication of the
accuracy of the vacuum-saturation approximation. In the case of ${\cal
H}_8$, this ratio yields ${\cal H}_8/{\cal H}_1^{\rm VS}$ in the limit
$T,T' \rightarrow \infty$. We obtain values for ${\cal G}_1$ and ${\cal
H}_1$ in the vacuum-saturation approximation from the relations
\begin{equation}
\sum_x\langle 0|\chi^\dag(x,T)\psi(x,T)\psi^\dag(0,0)\chi(0,0)|0\rangle
\stackrel{T \rightarrow \infty}{\longrightarrow} {\cal G}_1^{\rm VS}
{\rm exp}(-E_S T)
\label{eqn:g1v}
\end{equation}
and
\begin{equation}                                                             
\sum_x\langle 0|\chi^\dag(x,T)(-{\textstyle\frac{i}{2}}
\stackrel{\leftrightarrow}{\bf D})
\psi(x,T)\cdot \psi^\dag(0,0)(-{\textstyle\frac{i}{2}}
\stackrel{\leftrightarrow}{\bf D})
\chi(0,0)|0\rangle 
\stackrel{T \rightarrow \infty}{\longrightarrow} {\cal H}_1^{\rm VS}
{\rm exp}(-E_P T)\,,
\end{equation}                                                             
which follow from the fact that only the lowest-lying intermediate state
with the correct quantum numbers contributes to the amplitude in the
limit $T\rightarrow \infty$. Note that we can write
\begin{equation}
{\cal G}_1^{\rm VS}= a_p^2\,,
\label{a_P}
\end{equation}
where one factor of $a_p$ is from the point source and the other is from
the point sink. If we replace the point source by another (extended)
source, the coefficient of the exponential is of the form $a_p a_x$,
while if we use this new extended-source operator for both source and
sink, the coefficient is $a_x^2$. Thus, introducing an extended source
which has a greater overlap with the ground state gives us an
alternative method of extracting $a_p$ and, hence, ${\cal G}_1$. Similar
comments hold for ${\cal H}_1$ and ${\cal F}_1$. We calculate the ${\cal
F}_1$'s from
\begin{equation}
\label{eqn:f1/g1}
{-\sum_x\langle 0|\chi(x,T)^\dag\Delta^{(2)}(*)\psi(x,T) S(0) |0\rangle 
\over \sum_x\langle 0|\chi(x,T)^\dag\psi(x,T) S(0) |0\rangle}
\stackrel{T \rightarrow \infty}{\longrightarrow} {{\cal F}_1^{\rm VS}
\over {\cal G}_1^{\rm VS}}\,,
\end{equation}
where $\Delta^{(2)}(*)$ denotes any of the discretizations of ${\bf
D}^2$ in Eq.~(\ref{disc-delsq}), and $S(0)$ is any source with a finite
overlap with the lowest $S$-wave state on time slice $0$.

In order to evaluate these matrix elements, we must calculate
bottom-quark propagators $G(x;y)$ on the lattice. Following Lepage {\it
et al.}\ \cite{lepage-et-al}, we calculate the retarded propagator
$G_r({\bf x},t;0)$ by iterating the equation
\begin{equation}
\label{eqn:g}
G_r({\bf x},x_0+1;0)=(1-H_0/2n)^n U_{{\bf x},x_0}^{\dag} 
(1-H_0/2n)^n G_r({\bf x},x_0;0)+ \delta_{\bf x,0}\delta_{x_0+1,0}\,,
\label{G_r}
\end{equation}
setting $G({\bf x},x_0;0)=0$ for $x_0 < 0$. In Eq.~(\ref{G_r}), $H_0 =
-\Delta^{(2)}/2 M_0 - h_0$, $\Delta^{(2)}$ is the gauge-covariant
discrete Laplacian, which is given by the expression in
Eq.~(\ref{cov-2}) with $u_0$ set to unity, $h_0=3(1-u_0)/M_0$, and $M_0$
is the bare bottom-quark mass. We note that our bare bottom-quark mass
is defined to be $u_0$ times that of Ref.~\cite{nrqcd}. The value two
for the discretization parameter $n$ turns out to be adequate for our
calculations.

An expression that is similar to Eq.~(\ref{G_r}) exists for the advanced
propagator $G_a$. The relation $G_r(x;y)=G_a^\dag(y;x)$ makes it
possible to rewrite amplitudes, interchanging sources and sinks. Such a
rewriting allows one to start all propagator calculations from a noisy
(point or extended) source, rather than a point source and, thereby, to
reduce both the statistical error and the number of calculational steps.

\section{The relationship between lattice and continuum matrix elements}
\label{sec:pert-coeffs}

We wish to relate our lattice results to the continuum [modified minimal
subtraction scheme ($\overline{\rm MS}$)] matrix elements that are used
in phenomenology. Lattice matrix elements and continuum matrix elements
differ only in the choice of ultraviolet regulator. Furthermore, a
change of ultraviolet regulator is dependent only on the large-momentum
(short-distance) parts of an amplitude. Consequently, asymptotic freedom
allows us to compute the short-distance coefficients that relate the
lattice matrix elements to the continuum matrix elements in a
perturbation series in the strong coupling $\alpha_s$. The
short-distance coefficients are independent of the hadronic state.
Therefore, for purposes of computing the short-distance coefficients, we
choose, for convenience, to evaluate the operators in free $Q\bar Q$
states.

We can expand the lattice-regulated matrix element of an operator in
terms of continuum-regulated matrix elements of a complete set of
operators:
\begin{equation}
\langle {\cal O}_i\rangle_L=\sum_j c_{ij}\langle {\cal O}_j\rangle_C\,,
\label{op-exp}
\end{equation}
where the $c_{ij}$ are the short-distance coefficients, $\langle{\cal
O}\rangle$ is the matrix element of the operator ${\cal O}$ in a
free $Q\bar Q$ state, and the subscripts $L$ and $C$ indicate the
lattice- and continuum-regulated matrix elements, respectively. The
matrix elements and short-distance coefficients can be expanded in
perturbation series:
\begin{mathletters}
\begin{eqnarray}
\langle {\cal O}_i\rangle_L&&=\langle {\cal O}_i\rangle_L^{(0)}
+\alpha_s\langle {\cal O}_i\rangle_L^{(1)}+\cdots,\\
\langle {\cal O}_i\rangle_C&&=\langle {\cal O}_i\rangle_C^{(0)}
+\alpha_s\langle {\cal O}_i\rangle_C^{(1)}+\cdots,\\
c_{ij}&&=c_{ij}^{(0)}+\alpha_s c_{ij}^{(1)}+\cdots.
\end{eqnarray}
\label{pert-series}%
\end{mathletters}%
For simplicity, we use the same definition of $\alpha_s$ and the same
scale for $\alpha_s$ in all three expansions in Eq.~(\ref{pert-series}).

At zeroth order in the perturbation series, the momentum-space
expression for a lattice operator is equal to the momentum-space
expression for the corresponding continuum operator, plus terms of
higher order in the lattice spacing $a$ times the momenta. Therefore,
\begin{equation}
c_{ii}^{(0)}=1\,,
\label{leading-coeff}
\end{equation}
and
\begin{equation}
c_{ij}^{(0)}=0 \hbox{ for } 
{\rm Dim}\,{\cal O}_j< {\rm Dim}\,{\cal O}_i\,,
\end{equation}
where ${\rm Dim}\,{\cal O}$ is the mass dimension (or, equivalently,
order in $v$) of the operator ${\cal O}$. For the operators that we
consider in this paper,
\begin{equation}
c_{ij}^{(0)}=0 \hbox{ for } i\neq j.
\label{leading-coeff-mix}
\end{equation}

Since our lattice NRQCD action is accurate only to leading order in
$v$, only the following mixings can be treated consistently:
${\cal G}_{1L}$ into ${\cal G}_{1C}$
and ${\cal F}_{1C}$, ${\cal F}_{1L}$ into ${\cal F}_{1C}$,
${\cal H}_{1L}$ into ${\cal H}_{1C}$ and ${\cal H}_{8C}$, and
${\cal H}_{8L}$ into ${\cal H}_{8C}$ and ${\cal H}_{1C}$.
Therefore, we need consider, at most, two operators in the expansion
(\ref{op-exp}). Then, using Eqs.~(\ref{pert-series}),
(\ref{leading-coeff}), and (\ref{leading-coeff-mix}), we equate the
terms of order $\alpha_s^1$ in Eq.~(\ref{op-exp}) to obtain
\begin{equation}
\langle {\cal O}_i\rangle_L^{(1)}-\langle {\cal O}_i\rangle_C^{(1)}=
c_{ii}^{(1)}\langle {\cal O}_i\rangle_C^{(0)}
+c_{ij}^{(1)}\langle {\cal O}_{j}\rangle_C^{(0)} \hbox{ for } i\neq j,
\label{order-1}
\end{equation}
where no sum over $j$ is implied. The quantities on the left side of
Eq.~(\ref{order-1}) are computed in perturbation theory. We determine
the short-distance coefficients on the right side of Eq.~(\ref{order-1})
by expanding the quantity on the left side of Eq.~(\ref{order-1}) in
powers of the external $Q\bar Q$ 3-momenta and by choosing free $Q\bar
Q$ states with particular color (and, in general, spin) quantum numbers.

In the expansion of the quantity on the left side of Eq.~(\ref{order-1})
in powers of the external $Q\bar Q$ 3-momenta, the various terms are
infrared finite, to the extent that the behavior of the integrand in the
lattice matrix element matches the behavior of the integrand in the
continuum matrix element at small loop momentum. The expansions for the
various mixings that we have mentioned above yield, at most, a linear
infrared divergence in the lattice and continuum matrix elements. Since
our lattice action (and, implicitly, our continuum action) are
accurate to leading order in $v^2$, those divergences cancel between the
lattice and continuum matrix elements on the left side of
Eq.~(\ref{order-1}).

In general, infrared divergences in differences between lattice and
continuum matrix elements cancel, provided that one works consistently
to a given order in $v$. This means that, in order to compute
coefficient of the mixing of a lattice matrix element into a continuum
matrix element of relative order $v^n$, one must employ lattice and
continuum actions that are accurate to relative order $v^n$. Then,
the small-loop-momentum behaviors of the lattice and continuum
contributions on the left side of Eq.~(\ref{order-1}) will be the same,
and infrared divergences in the mixing coefficient will cancel. On the
other hand, one should not compute operator mixings that exceed the
accuracy in $v$ of the action. For example, since we use actions
of leading order in $v$, we do not compute the one-loop correction to
the mixing of ${\cal G}_{1L}$ into ${\cal F}_{1C}$ (relative order
$v^2$). If one were to carry out such a computation, using the
leading-order NRQCD lattice and continuum actions, then the
expression for the one-loop correction would contain a cubic leading
infrared divergence in both the lattice and continuum contributions on
the left side of Eq.~(\ref{order-1}). The cubic leading divergence would
cancel, but, owing to the absence of order-$v^2$ terms in the
action, a linear subleading divergence would persist.

Applying Eqs.~(\ref{op-exp}), (\ref{leading-coeff}), and
(\ref{leading-coeff-mix}) to the operator matrix elements that we
consider in this paper, we obtain
\begin{mathletters}
\begin{eqnarray}
{\cal G}_{1L} &=& (1+\epsilon) {\cal G}_1\,,\\
{\cal F}_{1L} &=& (1+\gamma ) {\cal F}_1 + \phi {\cal 
G}_1\,,\label{eqn:s-wave-b}
\end{eqnarray}
\label{eqn:s-wave}%
\end{mathletters}%
and
\begin{mathletters}
\begin{eqnarray}
{\cal H}_{1L} &=& (1+\iota ) {\cal H}_1 + \kappa {\cal H}_8, \\
{\cal H}_{8L} &=& (1+\eta) {\cal H}_8 + \zeta {\cal H}_1,
\end{eqnarray}
\label{eqn:p-wave}%
\end{mathletters}%
where we have dropped the subscript $C$ on the continuum matrix
elements, and the coefficients $\epsilon$, $\gamma$, $\phi$, $\iota$,
$\kappa$, $\eta$ and $\zeta$ are of order $\alpha_s$. It turns out, in
an explicit calculation, that the coefficient $\kappa$ actually
vanishes in order $\alpha_s$. Details of the calculations of
these coefficients will be given elsewhere \cite{bks-6}.

We note that the perturbation series for the $\overline{\rm MS}$
continuum short-distance coefficients that relate matrix elements to
physical quantities contain renormalon ambiguities. The $\overline{\rm
MS}$ continuum operator matrix elements contain compensating ambiguities,
and, so, the physical quantities are ambiguity free \cite{bodwin-chen}.
In contrast, the lattice operator matrix elements and the short-distance
coefficients that relate them to physical quantities are free of
renormalon ambiguities \cite{bodwin-chen}. Consequently, the
perturbation series that relate the lattice and the $\overline{\rm MS}$
continuum operator matrix elements contain renormalon ambiguities. At
the one-loop order to which we work, the factorial growth of the series
associated with the presence of renormalons is unimportant. However,
because the series that relate the lattice and the $\overline{\rm MS}$
continuum operator matrix elements (and the series that relate physical
quantities and the $\overline{\rm MS}$ continuum operator matrix
elements) ultimately fail to converge, the value of an $\overline{\rm
MS}$ continuum matrix element is meaningful only if one specifies the
order in perturbation theory that is employed in computing it.

\section{Results}

\subsection{Lattice Computation of the Matrix Elements}

For the lattice calculations, we use gauge configurations generated by
the HEMCGC Collaboration \cite{hemcgc} with two flavors of light
dynamical staggered quarks on a $16^3 \times 32$ lattice at $\beta
\equiv 6/g^2 = 5.6$. We use all $399$ configurations with light-quark
mass $m=0.01$ (in lattice units) and $200$ configurations with quark
mass $m=0.025$. As we have already mentioned, we follow Lepage {\it et
al.} in choosing $u_0 = [\frac{1}{3}{\rm Tr}U_{\rm plaq}]^{1/4}$ as
our definition of the tadpole contribution to $U$. Our measurements yield
$u_0=0.866985(11)$ at $m=0.01$ and $u_0=0.866773(12)$ at $m=0.025$.
Since these are so close, we use $0.866859$ for our perturbative
calculations. We choose our bare bottom-quark mass to be $M_b = 1.56
\approx 1.80 u_0$, where $1.80$ is the value chosen by the NRQCD
Collaboration \cite{nrqcd} to yield the best fit to the
$\Upsilon$-$\chi_b$ and $\Upsilon$-$\Upsilon'$ mass splittings.

\begin{figure}[htb]
\epsfxsize=6in
\centerline{\epsffile{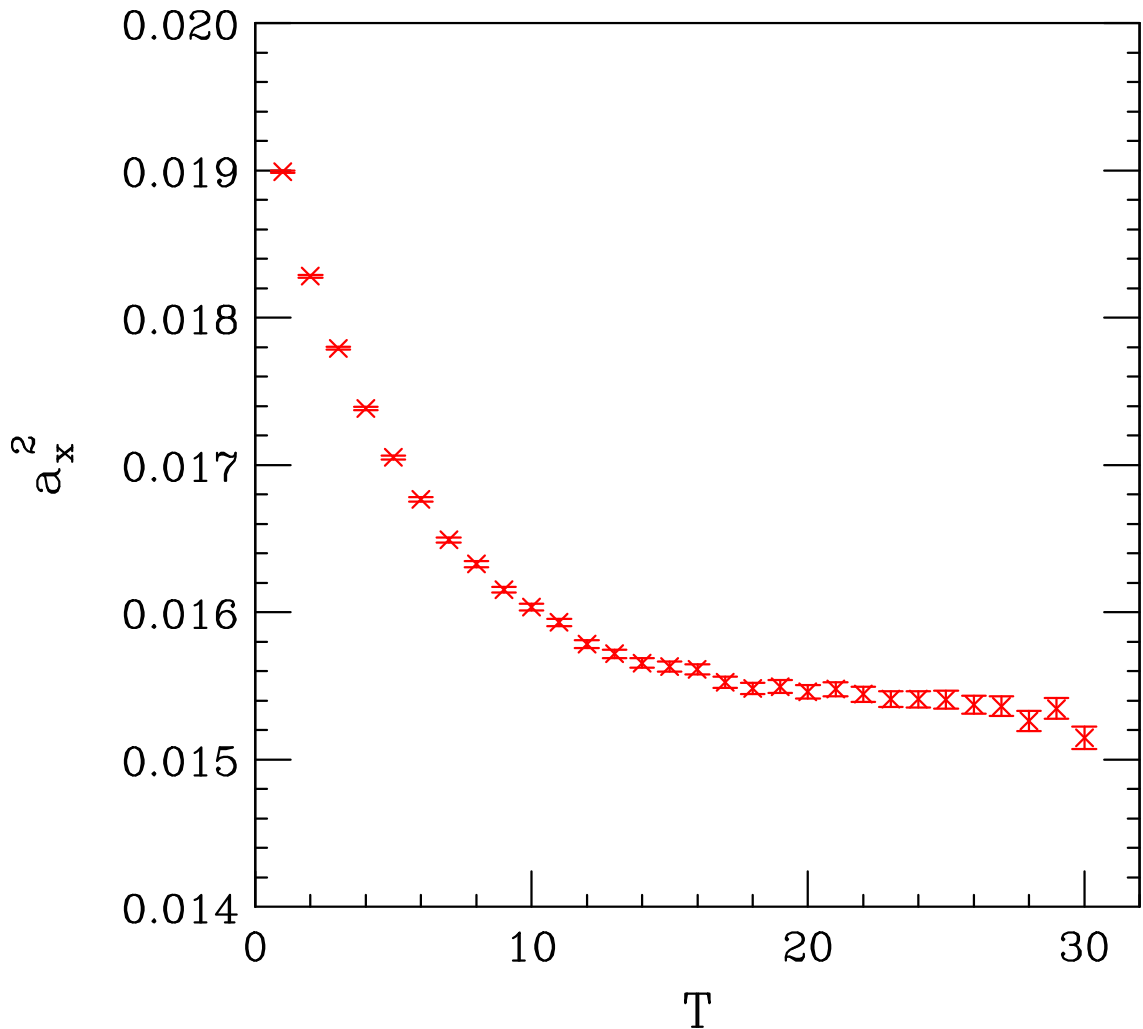}}
\caption{Effective $a_x^2$ as a function of $T$ for $S$-wave bottomonium.}
\label{fig:xwave}
\end{figure}

\begin{figure}[htb]                                                  
\epsfxsize=6in                                                    
\centerline{\epsffile{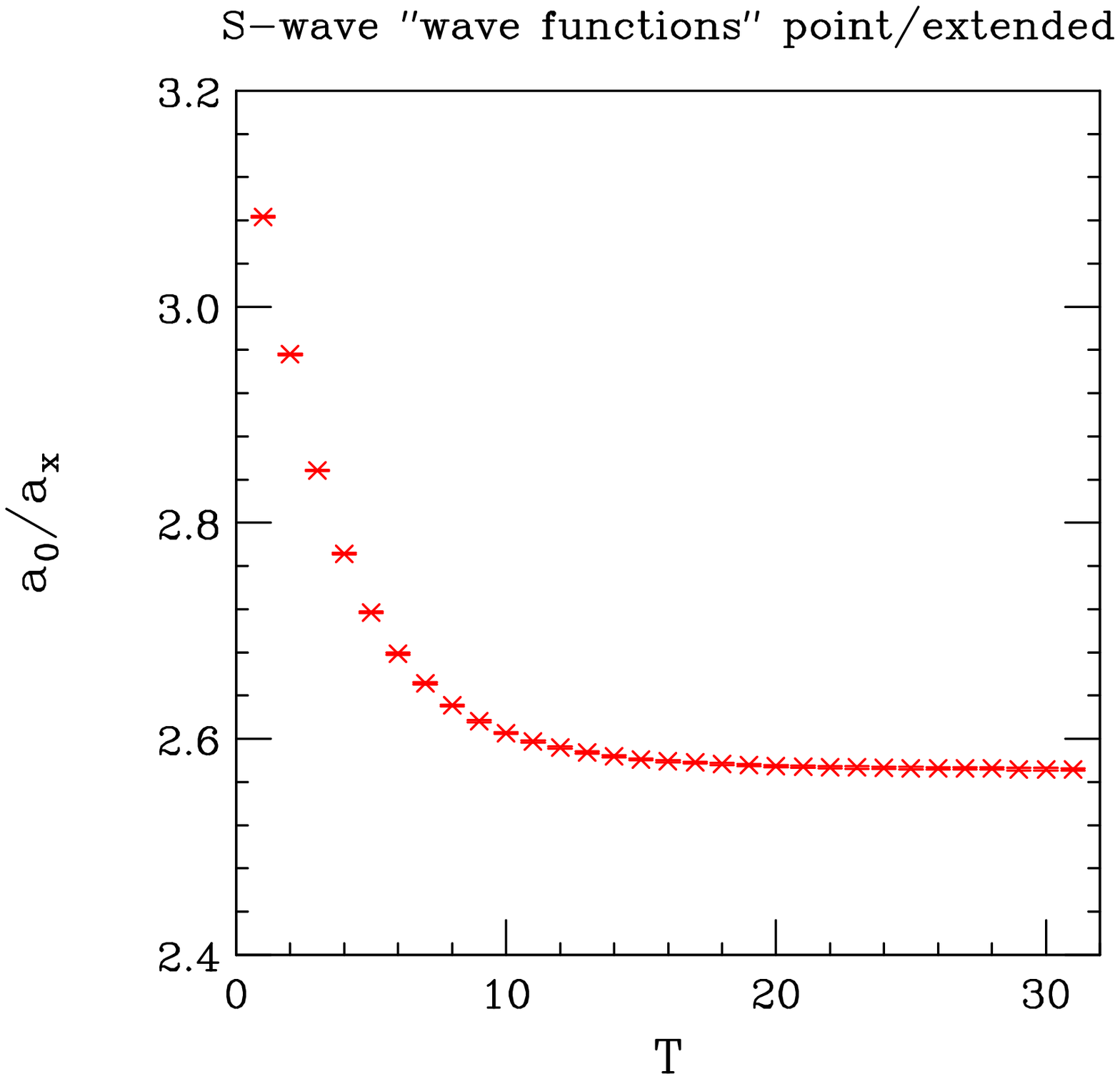}}                        
\caption{Ratio of the extended-point propagator to the extended-extended
propagator as a function of $T$ for $S$-wave bottomonium.}
\label{fig:ratio}                                                     
\end{figure}

To calculate the required matrix elements, we first gauge fix our
configurations to the Coulomb gauge. We then generate the advanced and
retarded bottom-quark propagators from a stochastic estimator to an
$S$-wave point source, a stochastic estimator to an $S$-wave Gaussian
source, and a stochastic estimator to a $P$-wave point source for each
color on each time slice. The width of the Gaussian source is chosen to
be $2.5$ in lattice units, which is approximately the radius of
$\Upsilon$ or $\eta_b$. From these we calculate the $S$- and $P$-wave
bottomonium propagators, with both point and Gaussian sources and sinks,
and the matrix elements of Fig.~\ref{fig:matrix} and
Eq.~(\ref{eqn:f1/g1}). Because the extended source has a larger overlap
with the ground state than does the point source, we extract $a_x^2$
from fits of the propagator with an extended (Gaussian) source and sink
to the form $a_x^2 {\rm exp}(-E T)$ for large $T$. We then calculate the
ratio $a_p/a_x$ from fits of the ratio of the propagator with extended
source and point sink to the propagator with extended source and sink.
Finally, we extract ${\cal G}_1({\cal H}_1)=2 a_0^2$. [An extra factor
of two appears here relative to Eq.~(\ref{a_P}) because, owing to the
spin independence of the lattice action at leading order in $v$, we
compute propagators for only a single spin component.] In the case of
${\cal G}_1$, the direct extraction from the point-source-point-sink
propagator gives a result that is consistent with this indirect method.
However, for ${\cal H}_1$ the point-point propagator is very noisy and
shows no sign of a plateau in the effective wave-function plot. In this
case the indirect method is required. Figure~\ref{fig:xwave} shows the
effective wave function as a function of $T$ for the $S$-wave
extended-extended propagator. Figure~\ref{fig:ratio} shows the ratio of
the $S$-wave extended-point propagator to the $S$-wave extended-extended
propagator as a function of $T$.


Our estimates of ${\cal F}^{\rm VS}_1/{\cal G}^{\rm VS}_1$ from the
various discretizations of ${\bf D}^2$ are obtained from fits to the
propagator ratios of Eq.~(\ref{eqn:f1/g1}) for the extended source. Our
point-source results for these ratios are completely consistent with the
extended-source results.

\begin{figure}[htb]
\epsfxsize=6in
\centerline{\epsffile{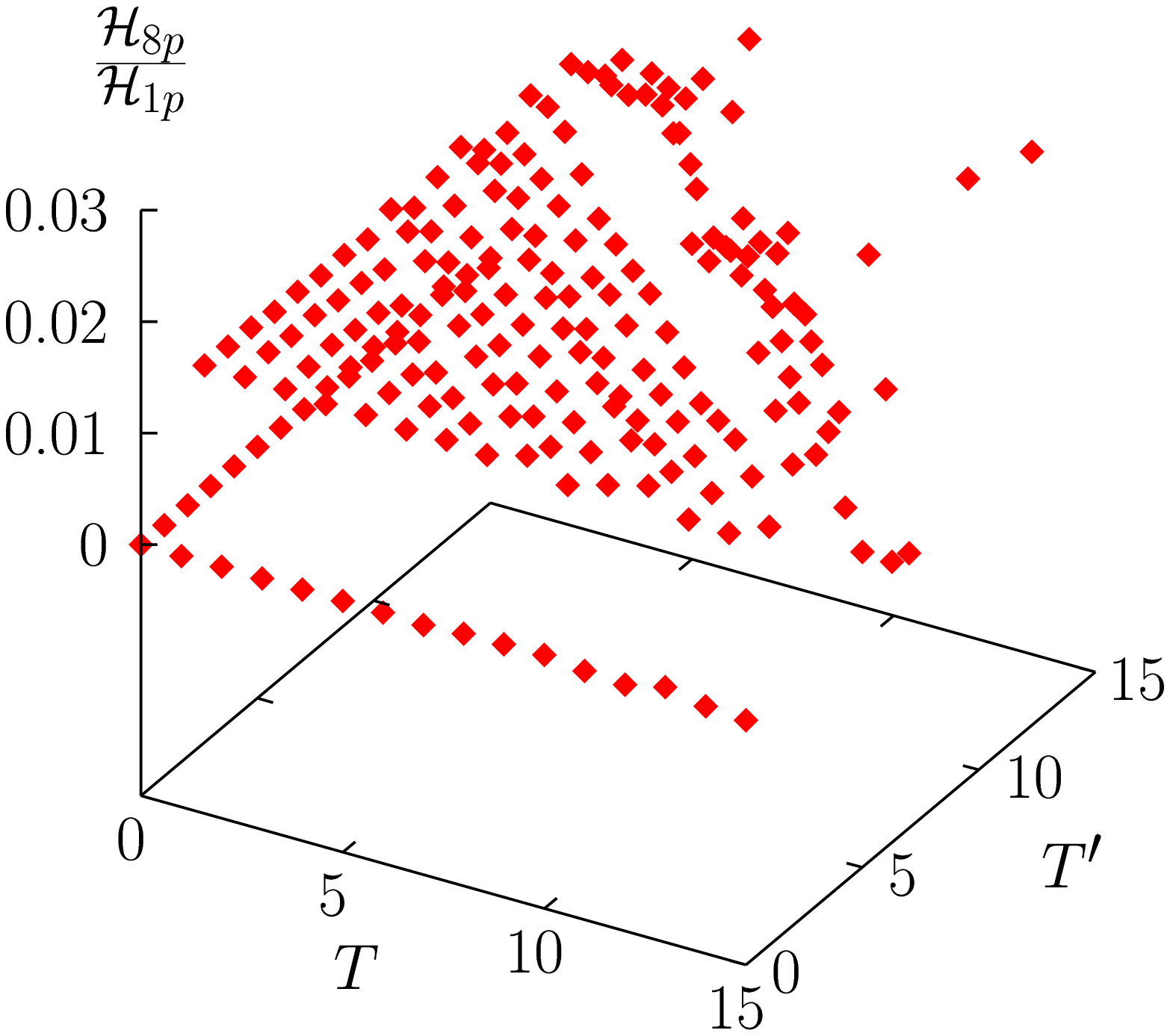}}
\caption{${\cal H}_{8L}/{\cal H}_{1L}$ as a function of $T$ and $T'$ for
point sources/sinks. Error bars have been suppressed to make the graph
more easily readable. The suffix $p$ indicates that we used a point
source and point sink.}
\label{fig:h8p}
\end{figure}

Finally, we extract the ratio ${\cal H}_8/{\cal H}^{\rm VS}_1$ from the
quantity represented in Fig.~\ref{fig:matrix}, where the 4-point vertex
denotes the octet operator of Eq.~(\ref{eqn:h8}). Its value for the case
of a point source and sink is plotted in Fig.~\ref{fig:h8p}. We consider
fits over the ranges $T_1 \leq T,T'\leq T_2$, for all choices of $T_1$
and $T_2$, excluding overlaps. From these we choose a ``best'' fit, {\it
i.e.}, one with a good confidence level, small error, and a reasonably
large range $T_2-T_1$. The chosen best fit is over the range 2--12 and
has a confidence level of 40\%. It yields a value ${\cal H}_{8L}/{\cal
H}_{1L}=0.01565(8)$. In comparison, the fit with the highest confidence
level (99.8\%) is over the range 6--8 and yields a value ${\cal
H}_{8L}/{\cal H}_{1L}=0.01540(16)$, which is in agreement the selected
fit. The results for the extended source are consistent the results for
a point source, but the plateau occurs roughly one unit later in $T,T'$,
and the ``data'' are noisier. We estimate the systematic error in ${\cal
H}_{8L}/{\cal H}_{1L}$ by examining the entire plateau, both for the
point-source data and for the extended-source data, and determining the
range of fluctuations in the region in which the signal-to-noise ratio
is appreciable.

Note that there is clear evidence for a plateau in ${\cal H}_{8L}/{\cal
H}_{1L}$ for $T,T' \gtrsim 1$, and, so, we are justified in assuming
that the asymptotic behavior occurs for relatively small $T,T'$, where
the signal-to-noise ratio is relatively good. An analysis of the
effective wave function for the $P$-wave state shows a plateau that
starts at $T\approx 10$. However, we observe in the ratios that we use to
calculate ${\cal F}_1$, in which we have decent signals out to $T=31$,
that the plateau can start much before effective masses and effective
wave functions indicate that one has obtained a pure state. This is also
the case for the data that we present later on $G_1/G^{\rm VS}_1$.
Presumably, the early onset of a plateau in these ratios indicates that
their values are not very different for the $1P$ and $2P$ states (and
$1S$ and $2S$ states).

The results for these lattice matrix elements are given in
Table~\ref{tab:mel}.
\begin{table}[htb]
\begin{tabular}{@{\hspace{0.3cm}}l@{\hspace{0.3cm}}@{\hspace{1.0cm}}c%
@{\hspace{1.0cm}}@{\hspace{1.0cm}}c@{\hspace{1.0cm}}}
                                     &      m=0.01       &      m=0.025  \\
\hline
${\cal G}_{1L}$                      &  $0.20479 \pm 0.00036 \pm 0.0028$  
                                     &  $0.21265 \pm 0.00061 \pm 0.0028$ \\
${\cal F}_{1L}({\rm non})/{\cal G}_{1L}$   &  $1.53074 \pm 0.00049 
\pm 0.001$   
                                     &  $1.54816 \pm 0.00069 \pm 0.0005$ \\
${\cal F}_{1L}({\rm cov})/{\cal G}_{1L}$   &  $0.99667 \pm 0.00037 
\pm 0.0005$  
                                     &  $1.00740 \pm 0.00049 \pm 0.0005$ \\
${\cal F}_{1L}({\rm non}_2)/{\cal G}_{1L}$ &  $0.86310 \pm 0.00027 
\pm 0.0005$  
                                     &  $0.87226 \pm 0.00037 \pm 0.0003$ \\
${\cal F}_{1L}({\rm cov_2})/{\cal G}_{1L}$ &  $1.23209 \pm 0.00022 
\pm 0.001$    
                                     &  $1.23961 \pm 0.00029 \pm 0.0005$ \\
${\cal H}_{1L}$                      &  $0.02016 \pm 0.00078 \pm 0.0014$  
                                     &          ----                     \\
${\cal H}_{8L}/{\cal H}_{1L}$        &  $0.01565 \pm 0.00008 \pm 0.001$   
                                     &          ----                         
\end{tabular}
\caption{Lattice bottomonium decay matrix elements for light quark masses
$m=0.01$ and $m=0.025.$}
\label{tab:mel}
\end{table}
The first error bar is statistical. The second error bar is a
combination of our estimate of the systematic error that arises from our
choice of fits and our estimate of the uncertainty that arises from the
fact that the propagators have not reached their asymptotic forms in the
region of measurement. We note that the dependence on the
light-quark mass is weak. For this reason and for the reason that we
have fewer configurations at the higher light-quark mass, we have not
calculated the $P$-wave matrix elements at $m=0.025$.

\subsection{Lattice-to-Continuum Conversion}

First, let us present the one-loop results for the coefficients that
relate the lattice matrix elements to their continuum ($\overline{\rm
MS}$) counterparts. These coefficients were defined and the method for
their calculation was outlined in the Sec.~\ref{sec:pert-coeffs}. The
loop integrals were evaluated numerically, using the adaptive Monte
Carlo routine VEGAS \cite{vegas}. The values of the coefficients, in
lattice units ($a=1$), are presented in Table~\ref{tab:coeff}. These
values depend on the value of the bottom-quark mass in lattice units.
However, as we have already discussed, we take the bottom-quark mass, in
lattice units, to be the same at both the values of the light-quark mass
$m$ that we use. Then, with the exception of $\zeta$, the coefficients
in Table~\ref{tab:coeff} depend on the light-quark mass only through the
scale of $\alpha_s$, which is proportional to $a$, since $a$ depends
(weakly) on $m$. $\zeta$ has additional dependence on $a$ and, hence, on
$m$, since it contains a term that is proportional to ${\rm ln}(\mu a)$,
where $\mu$ is the NRQCD factorization scale. We take $\mu=4.3~{\rm
GeV}$, which is close to $M_b(\overline{\rm MS})$.

\begin{table}[htb]
\begin{tabular}{@{\hspace{0.5cm}}l@{\hspace{0.5cm}}%
@{\hspace{2cm}}c@{\hspace{2cm}}}
Coefficient            &     Value\\
\hline
$\epsilon$             &     -0.4387~$\alpha_s$      \\
$\gamma({\rm non})$          &     -0.9622~$\alpha_s$      \\
$\gamma({\rm cov})$          &     -2.543~$\alpha_s$       \\
$\gamma({\rm non}_2)$        &     -0.9489~$\alpha_s$      \\
$\phi({\rm non})$            &     4.822~$\alpha_s$         \\
$\phi({\rm cov})$            &     3.729~$\alpha_s$        \\
$\phi({\rm non}_2)$          &     3.078~$\alpha_s$        \\
$\iota$                &     -1.232~$\alpha_s$       \\
$\eta$                 &     0.05484~$\alpha_s$      \\ 
$\zeta(m=0.01)$        &     -0.01285~$\alpha_s$    \\
$\zeta(m=0.025)$       &     -0.01680~$\alpha_s$
\end{tabular}
\caption{Coefficients relating lattice and continuum matrix elements. The
different versions of $\gamma$ and $\phi$ relate to the different 
discretizations of ${\bf D}^2$.}
\label{tab:coeff}
\end{table}

We convert the lattice matrix elements to continuum matrix elements
using the formulae of Eqs.~(\ref{eqn:s-wave}) and (\ref{eqn:p-wave}).
Here, we choose $\alpha_s = \alpha_P(1/a)$, where $\alpha_P$ is defined
in Ref.~\cite{nrqcd}. To convert to physical units, we use
$a^{-1}=2.44~{\rm GeV}$ for $m=0.01$ and $a^{-1}=2.28~{\rm GeV}$ for
$m=0.025$, as determined by the NRQCD Collaboration \cite{nrqcd} from the
$\Upsilon$-$\chi_b$ mass splitting. Then, the required values of
$\alpha_s$ are $\alpha_P(2.44~{\rm GeV})=0.2941\pm 0.0070$ and
$\alpha_P(2.28~{\rm Gev})=0.3056\pm 0.0076$, respectively. Our continuum
matrix elements are given in Table~\ref{tab:mec}. The first two error
bars arise from the statistical and systematic uncertainties in the
lattice calculation. The third error bar is our estimate of the
uncertainty from uncalculated two-loop corrections to the coefficients
in Table~\ref{tab:coeff}. This uncertainty is estimated as the greater
of $\alpha_s$ times the one-loop contribution and $\alpha_s^2$ times the
tree-level coefficient. Clearly this is the dominant uncertainty. In the
case of ${\cal F}_1$, this uncertainty is magnified because the left
side of Eq.~(\ref{eqn:s-wave-b}) is very close in size to the second
term on the right side of Eq.~(\ref{eqn:s-wave-b}). Consequently, our
calculation of ${\cal F}_1$ is very imprecise. The lattice operator
matrix elements ${\cal F}_1({\rm cov})$, ${\cal F}_1({\rm non})$, and
${\cal F}_1({\rm non}_2)$ all yield values of ${\cal F}_1$ that are
consistent with zero. Furthermore, the error bars in each case are
larger than the differences between the central values. The operator
matrix element ${\cal F}_1({\rm non}_2)$ yields the smallest
uncertainties, and it is the value that derives from this matrix element
that we report in Table~\ref{tab:mec}.

There are some additional uncertainties that are not included in
Table~\ref{tab:mec}. One is the uncertainty that arises from the
uncertainty in the NRQCD Collaboration's determination of $a$
\cite{nrqcd}. We include only the statistical uncertainty in $a^{-1}$ in
our calculation. The NRQCD collaboration also reports an order-$a^2$
uncertainty and an order-$v^4$ uncertainty. The former is equivalent, in
NRQCD, to the order-$v^2$ uncertainty, which we estimate later. The
latter we ignore in comparison with the order-$v^2$ uncertainty in our
calculation. The uncertainty in $a$ translates into an uncertainty in
${\cal G}_1$ of approximately 8\% at $m=0.01$ and 19\% at $m=0.025$. In
the case of ${\cal H}_1$, it leads to an uncertainty of approximately
14\% at $m=0.01$. Another uncertainty arises from the neglect of
corrections of higher order in $v^2$ in the action. These are nominally
of order 10\%, but are expected to be much smaller for spin-averaged
quantities. (Of course, in order to obtain a spin-averaged value of
${\cal G}_1$, one would need to observe the $\eta_b$ and measure its
decay width into $\gamma\gamma$.)  Finally, there is the effect of
order-$a^2$ corrections in the gluon and light-quark sectors, which cause
appreciable flavor-symmetry breaking at the values of the lattice
spacing that we use. These effects could best be estimated by repeating
the calculation at a different value of $\beta$.

\subsection{Phenomenological Value of the Matrix Element}

We obtain a phenomenological estimate for ${\cal G}_1$ from the leptonic
decay width of $\Upsilon$ \cite{leptonic-width}
\begin{equation}
\Gamma(\Upsilon \rightarrow e^+ e^-)\approx 
{2\pi Q_b^2 \alpha^2 \over 3 M_b^2({\rm pole})}
\left(1-{16\alpha_s \over 3 \pi}\right){\cal G}_1\,.
\label{leptonic-width}
\end{equation}
Here, we use $M_b({\rm pole})=5.0\pm 0.2~{\rm GeV}$ \cite{Davies:1994pz},
$\alpha(M_b)=1/132$, $\alpha_s(M_b)=0.212$, and $\Gamma(\Upsilon
\rightarrow e^+ e^-)=1.32\pm 0.05~{\rm Gev}$ \cite{pdg}. 
The value of ${\cal G}_1$ given in Table~\ref{tab:mec} includes only the
experimental uncertainty. 

In extracting the phenomenological value of ${\cal G}_1$, we have not
included the relative-order-$\alpha_s^2$ correction to $\Gamma(\Upsilon
\rightarrow e^+ e^-)$ \cite{beneke-et-al}. It would be inconsistent to
include this correction without also including the order-$\alpha_s^2$
corrections to the short-distance coefficients that relate the lattice
operator matrix elements to the continuum ones. The
relative-order-$\alpha_s^2$ correction to $\Gamma(\Upsilon \rightarrow
e^+ e^-)$ contains a large dependence on the NRQCD factorization scale
$\mu$. If we did include this correction in our extraction, then the
phenomenological value of ${\cal G}_1$ would range from $3.76~{\rm
GeV}^3$ to $8.77~{\rm GeV}^3$ as the $\mu$ ranges from $1~{\rm GeV}$ to
$M_b$. This large $\mu$ dependence and the large size of the correction
at $\mu=M_b$ would seem to indicate that the uncertainty in the
phenomenological value may be close to 100\%.  However, experience with
one-loop corrections to quarkonium decay processes suggests that such
large corrections may be canceled by the order-$\alpha_s^2$ corrections
to the short-distance coefficients that relate the lattice operator
matrix elements to the continuum ones. Certainly, the large $\mu$
dependence in the relative-order-$\alpha_s^2$ correction to
$\Gamma(\Upsilon \rightarrow e^+ e^-)$ would be compensated exactly by a
large $\mu$ dependence in the order-$\alpha_s^2$ corrections to the
short-distance coefficients.

The uncertainty in the phenomenological  value of ${\cal G}_1$ that arises
from the uncertainty in the value of $M_b$ is about 8\%. This is
negligible in comparison with the uncertainty associated with the
perturbation expansion. Given present theoretical uncertainties, it is
not yet possible to extract ${\cal F}_1$ from experiment.

${\cal H}_1$ and ${\cal H}_8$ are related to the $\chi_b$ decay widths,
which have not yet been determined in experiments. Large corrections to
the perturbation series \cite{p-wave-decays} are likely to be important
sources of uncertainty in the determination of these quantities, once
experimental data become available.

\begin{table}[htb]
\begin{tabular}{@{\hspace{0.2cm}}l@{\hspace{0.8cm}}%
c@{\hspace{0.8cm}}c@{\hspace{0.8cm}}c@{\hspace{0.8cm}}}
             & \multicolumn{2}{c@{\hspace{1.6cm}}}{Calculation ($n_f=2)$} & \\
                         & Lattice Units       & Physical Units & 
Phenomenology \\
\hline
$m=0.01$                      &                     &            &           \\
${\cal G}_1$                  & 0.2351(4)(32)(240)  & 3.416(6)(47)(340)~GeV$^3$ 
                              & 3.86(14)~GeV$^3$                              \\
${\cal F}_1 / {\cal G}_1$     & -0.8 --- 0.3        & -5 --- 2~GeV$^2$
                              &      -----                                   \\
${\cal H}_1$                  & 0.032(1)(2)(5)      & 2.7(1)(2)(5)~GeV$^5$
                              &      -----                                   \\
${\cal H}_8 / {\cal H}_1$     & 0.01354(5)(63)(390) & 0.002275(9)(105)(660)%
~GeV$^{-2}$                   &      -----                                   \\
$m=0.025$                     &                     &            &           \\
${\cal G}_1$                  & 0.2456(7)(32)(270)  & 2.911(8)(38)(320)~GeV$^3$
                              & 3.86(14)~GeV$^3$                              \\
${\cal F}_1 / {\cal G}_1$     & -0.9 --- 0.3        & -4.7 --- 1.5~GeV$^2$
                              &      -----                                   
\end{tabular}
\caption{Continuum $\overline{\rm MS}$ bottomonium decay matrix elements
from our lattice calculations with two dynamical light quarks ($n_f=2$)
and, for comparison, a phenomenological value of ${\cal G}_1$. The error
bar on the phenomenological value of ${\cal G}_1$ does not include the
theoretical uncertainty.}
\label{tab:mec}
\end{table}

\subsection{Extrapolation to Physical Light-Quark Values}

We use linear extrapolation methods to estimate the calculated matrix
elements at the physical values of the light-quark masses and at the
physical number of light-quark flavors. Extrapolating to $m=0$, we find
that ${\cal G}_1=3.75(1)(8)(38)~{\rm GeV}^3$. To extrapolate to the
physical values of the light-quark masses, we use the HEMCGC light-hadron
spectroscopy measurements on the gauge configurations that we employ
\cite{hemcgc} to estimate that one-third the mass of the strange quark
is approximately $0.0071$, in lattice units. Then, we extrapolate ${\cal
G}_1$ to this value of $m$. Note that, since we are using linear
extrapolations in both $m$ and the number of flavors, this procedure
yields the same result as would setting $m_s=0.02$ and $m_u=m_d=0$.
Finally, we use our results for quenched QCD at $\beta=6.0$ \cite{bks-1}
to extrapolate to three light-quark flavors, obtaining ${\cal
G}_1=4.10(1)(9)(41)~{\rm GeV}^3$. This result is approximately 6\%
higher than the phenomenological  value. Similarly, extrapolations to
three light-quark flavors (with no extrapolation in $m$) yield ${\cal
H}_1 \approx 3.3~{\rm GeV}^5$ and ${\cal H}_8 / {\cal H}_1 \approx
0.0018~{\rm GeV}^{-2}$.

\subsection{Tests of the Vacuum-Saturation Approximation}

Our lattice calculations permit us to test the validity of the
vacuum-saturation approximation for ${\cal G}_1$ and ${\cal H}_1$. NRQCD
predicts that
\begin{eqnarray}
{\cal G}_1/{\cal G}_1^{\rm VS}& = & 1+{\cal 
O}(v^4)\label{vac-sat-ratio1}\\
{\cal H}_1/{\cal H}_1^{\rm VS}& = & 1+{\cal O}(v^4).
\label{vac-sat-ratio2}
\end{eqnarray}
Note that, although our lattice action is accurate only to leading order
in $v$, it {\it does} contain interactions of relative order $v^2$,
which arise through the terms proportional to the gauge field in the
covariant derivatives. These interactions allow for the spin-independent
emission of transverse gluons, which produces the leading correction to
the vacuum-saturation approximation.\footnote{The inclusion of covariant
derivatives does not, of course, generate all of the
relative-order-$v^2$ contributions to the matrix elements we consider.
In particular, the action that we use contains none of the
spin-dependent terms that distinguish states with the same orbital
angular momentum but with different total angular momentum (such as the
$\Upsilon$ and the $\eta_b$).}

\begin{figure}[htb]                                                            
\epsfxsize=6in                                                              
\centerline{\epsffile{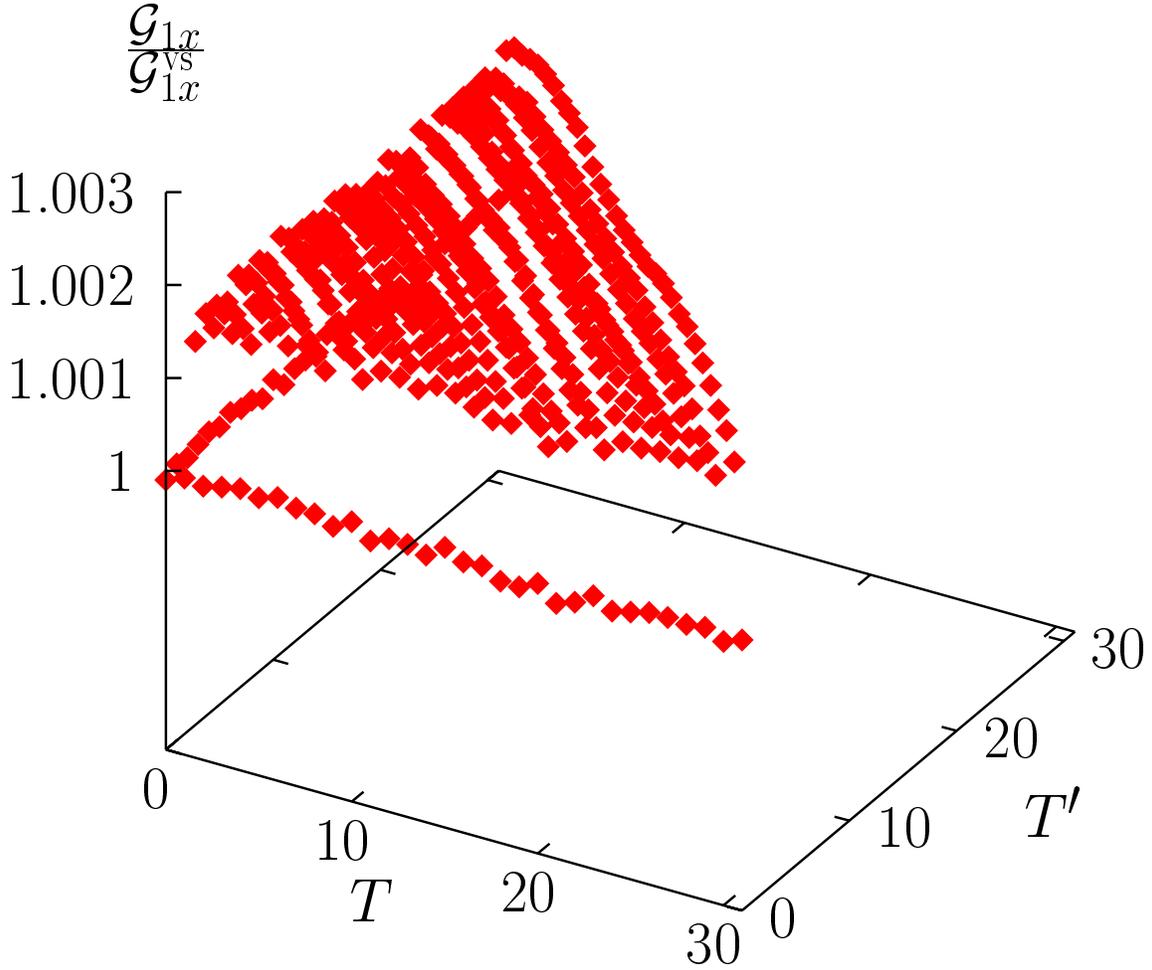}}
\caption{The ratio ${\cal G}_{1L}/{\cal G}_{1L}^{\rm VS}$ as a function
of $T$ and $T'$, where the subscript $x$ indicates the use of an
extended source. The error bars have been suppressed for clarity of
presentation.} 
\label{fig:g1}
\end{figure}

In order to test the relations (\ref{vac-sat-ratio1}) and
(\ref{vac-sat-ratio2}), we need to observe a plateau in ratios of the
form of Fig.~\ref{fig:matrix}. We have measured these ratios with both
point and extended sources. The detailed method of analysis is similar
to that for the ratio ${\cal H}_{8L}/{\cal H}_{1L}$ described above.
Because ${\cal G}_1 \propto {\cal G}_{1L}$ and ${\cal H}_1 \propto {\cal
H}_{1L}$, up to corrections of relative order $\alpha_s$, we use the
lattice quantities to evaluate the ratios in Eqs.~(\ref{vac-sat-ratio1})
and (\ref{vac-sat-ratio2}). We find that
\begin{eqnarray}                                                            
{\cal G}_1/{\cal G}_1^{\rm VS}& = & 1.0017(1) \\
{\cal H}_1/{\cal H}_1^{\rm VS}& = & 1.0049(2)\,.
\end{eqnarray}                                                                 
These results are consistent with $v^2$ being of order $0.1$ and justify
our use of the vacuum-saturation approximation in computing matrix
elements. Figure~\ref{fig:g1} shows the plateau in the ratio of lattice
matrix elements ${\cal G}_{1L}/{\cal G}_{1L}^{\rm VS}$. We note that the
plateau is reached for $T,T'
\gtrsim 1$. 

\subsection{The Nonrelativistic Energy}

For comparison with the work of the NRQCD Collaboration, we give our
estimate for their ``energy'' $E_{NR}$ \cite{Davies:1994mp}, which is
related to the $E_S$ of Eq.~(\ref{eqn:g1v}) by
\begin{equation}
E_{NR} = E_S + 2\, \ln u_0\,.
\end{equation}
For $m=0.01$ we obtain $E_{NR} = 0.4841(2)$, and for $m=0.025$ we obtain
$E_{NR} = 0.4901(3)$ (statistical errors only). This is to be compared
with the value $E_{NR} = 0.493(1)$ ($m$ unspecified) that was obtained
by the NRQCD Collaboration \cite{Davies:1994pz} using an action that is
accurate through next-to-leading order in $v^2$.

\section{Discussion and conclusions}

We have measured, in lattice simulations with two light-quark flavors,
the matrix elements of leading order and next-to-leading order in $v^2$
that mediate the decays of the $\Upsilon$ and the color-singlet and
color-octet matrix elements of leading order in $v^2$ that mediate the
decays of the $\chi_b$ states. We have also computed the relations
between the lattice matrix elements and continuum ($\overline{\rm MS}$)
matrix elements to one-loop accuracy and have converted our lattice
results to continuum results. We have extrapolated these results to the
case of three light-quark flavors.

Our measured value of the lowest-order $\Upsilon$-decay matrix element,
extrapolated to three flavors, is ${\cal G}_1=4.10(1)(9)(41)~{\rm
GeV}^3$. This is in good agreement with the phenomenological value
${\cal G}_1=3.86(14)$, which we extracted from the experimental value
for the leptonic decay width by using the perturbative-QCD expression
for the leptonic decay width, accurate through relative order
$\alpha_s$. This contrasts with the value for ${\cal G}$ that we
obtained in the quenched approximation \cite{bks-1}, which is 40--45\%
lower than the phenomenological value. The large size and large scale
dependence of the order-$\alpha_s^2$ correction to the leptonic width
\cite{beneke-et-al} suggest that the theoretical uncertainty in the
phenomenological value of the matrix element may be quite large.
However, one can utilize the order-$\alpha_s^2$ correction consistently
only when one has incorporated two-loop corrections into the
coefficients for the lattice-to-continuum conversion. Then the scale
dependence would be compensated exactly, and some of the large
correction would likely be canceled.

At present, the most that we can say about the next-to-leading-order
$\Upsilon$-decay matrix element ${\cal F}_1$ is that the continuum
($\overline{\rm MS}$) matrix element is probably negative. Notice that,
although the lattice matrix element is strictly positive, the
subtractions involved in converting it to a continuum matrix element
can, and apparently do, change its sign. If this result is maintained at
higher orders in perturbation theory, then it is clear that any simple
potential model, which must, of necessity, give a positive result,
cannot yield the correct values for such higher-order matrix elements.

We find that the value of the color-singlet $P$-wave matrix element
${\cal H}_1$ for three light-quark flavors is roughly 70\% higher than
the quenched value. The color-octet contribution to the $P$-wave decays,
which is mediated by ${\cal H}_8$, arises from a distinctive QCD effect:
the process in which the $b\bar{b}$ color-singlet $P$-wave state
fluctuates into a $b\bar{b}$ color-octet $S$-wave state plus a soft
gluon. Such a contribution is absent in simple potential models. We find
that the value of ${\cal H}_8$ for three light-quark flavors is about
40\% larger than the quenched value, but that the value of the ratio
${\cal H}_8/{\cal H}_1$ for three light-quark flavors is approximately
14\% lower than the quenched  value. According to the velocity-scaling
rules, both ${\cal H}_1$ and ${\cal H}_8$ are of order $v^4$
(Ref.~\cite{bbl}). Therefore, we expect $M_b^2{\cal H}_8/{\cal H}_1$ to
be of order unity times a factor $1/(2N_c)$, where $N_c=3$ is the number
of QCD colors, to account for the spin and color traces in the
definition of ${\cal H}_1$ (Ref.~\cite{Petrelli:1998ge}). Our result for
$M_b^2{\cal H}_8/{\cal H}_1$ is smaller than this estimate by about a
factor of three. Our values for ${\cal H}_1$ and ${\cal H}_8$ could be
used to make predictions for the, as yet, unmeasured $\chi_b$ decay
rates. However, as we have mentioned, large next-to-leading-order
corrections in the perturbation series for those decay rates
\cite{p-wave-decays} suggest that further theoretical progress may be
necessary in order to achieve a precise comparison with experiment.

Our results indicate that the quenched approximation yields a poor
estimate of the NRQCD matrix elements. As we have mentioned, the trends
in going from the quenched approximation to the physical number of
light-quark flavors can be understood in terms of a simple picture. The
lattice spacing and heavy-quark mass are determined by fitting to
bottomonium spectroscopy, which probes the wave functions at distances of
order $1/(M_b v)$. On the other hand, the decay matrix elements sample
the wave function and its derivatives at much shorter distances, of order
$a$. In the absence of the sea of light quark-antiquark pairs, the
strong coupling constant runs too fast, becoming too small at the
shorter distance. This leads to an underestimate of the decay matrix
elements, since the values of the wave function and its derivatives
depend on the strength of the potential at short distances. We note that
the $S$-wave matrix elements $G_1$ and $H_8$ both increase by about the
same fraction in going from the quenched approximation to three
light-quark flavors, while the $P$-wave matrix element $H_1$ increases
by a larger fraction. This may be because $P$-wave matrix elements
depend on the derivative of the wave function at the origin, as opposed
to the wave function at the origin, and, so, are more sensitive to
changes in the strength of the potential at short distances.

In order to improve the lattice estimates of the matrix elements for $S$-
and $P$-wave bottomonium decay, one would first need more precise
measurements of the lattice spacing $a$. A more stringent comparison
with experiment would also require a more precise determination of the
bottom-quark mass $M_b$, as well as a calculation to two-loop accuracy
of the perturbative coefficients that relate the lattice matrix elements
to the continuum matrix elements.\footnote{Alternatively, one could
carry out a non-perturbative (lattice) calculation of these coefficients
or compute the short-distance decay coefficients in lattice perturbation
theory, rather than in continuum perturbation theory.} In the case of
${\cal F}_1$ this last improvement is essential to obtain a useful
prediction of the continuum matrix element. Beyond this it would be
valuable to use gauge-field configurations that have been generated with
improved actions. Only when this has been done could one justify using
NRQCD actions that have been improved to higher orders in $v$ and $a$
for the extraction of bottomonium decay matrix elements.

\section{Acknowledgements}

We wish to thank the NRQCD Collaboration, and especially G.~P.~Lepage,
for discussions of their published and unpublished results. We thank
K.~Hornbostel for the use of his 3-loop program for evolving $\alpha_s$.
We also thank the HEMCGC Collaboration and, in particular, U.~M.~Heller
for giving us access to their gauge-field configurations. The lattice
calculations were performed on the Cray J90's at NERSC. Work in the High
Energy Physics Division at Argonne National Laboratory is supported by
the U.~S.~Department of Energy, Division of High Energy Physics, under
Contract No.~W-31-109-ENG-38. S.~K.\ is supported by Korea Research
Foundation Grant KRF-2001-015-DP0088.

\end{document}